\documentclass[%
aip,
apl,
 amsmath,amssymb,
 reprint,
 groupedaddress,
 showpacs]{revtex4-1}
\usepackage{bm}
\usepackage{graphicx}
\usepackage{gensymb}
\usepackage{amsmath}
\usepackage{amssymb}
\usepackage{dcolumn}
\usepackage{ulem}

\begin{document}

\title{Work function of bulk-insulating topological insulator Bi$_{2-x}$Sb$_x$Te$_{3-y}$Se$_y$}
\author{Daichi Takane,$^1$ Seigo Souma,$^{2,3}$ Takafumi Sato,$^{1,3}$ Takashi Takahashi,$^{1,3}$ Kouji Segawa,$^4$ and Yoichi Ando$^5$}

\affiliation{$^1$Department of Physics, Tohoku University, Sendai 980-8578, Japan\\
$^2$WPI Research Center, Advanced Institute for Materials Research, Tohoku University, Sendai 980-8577, Japan\\
$^3$Center for Spintronics Research Network, Tohoku University, Sendai 980-8577, Japan\\
$^4$Department of Physics, Kyoto Sangyo University, Kyoto 603-8555, Japan\\
$^5$Institute of Physics II, University of Cologne, K$\ddot{o}$ln 50937, Germany\\
}

\date{\today}

\begin{abstract}
    Recent discovery of bulk insulating topological insulator (TI) Bi$_{2-x}$Sb$_x$Te$_{3-y}$Se$_y$ paved a pathway toward practical device application of TIs. For realizing TI-based devices, it is necessary to contact TIs with a metal. Since the band-bending at the interface dominates the character of devices, knowledge of TIs' work function is of essential importance. We have determined the compositional dependence of work function in Bi$_{2-x}$Sb$_x$Te$_{3-y}$Se$_y$ by high-resolution photoemission spectroscopy. The obtained work-function values (4.95-5.20 eV) show a systematic variation with the composition, well tracking the energy shift of the surface chemical potential seen by angle-resolved photoemission spectroscopy. The present result serves as a useful guide for developing TI-based electronic devices.
\end{abstract}


\maketitle

  Topological insulator (TI) realizes a novel quantum state of matter characterized by the metallic Dirac-cone surface state with a helical spin texture which traverses the bulk band gap \cite{AndoReview, SCZhangReview, HasanReview}. Exotic topological phenomena of TIs \cite{ME, ME2, AQHE} as well as their device applications largely rely on the bulk-insulating nature and the dominant Dirac transport at the surface. However, it is known that the bulk-insulating nature is hard to be achieved in most of known TIs mainly due to defects in the crystals \cite{AndoReview}. Tetradymite solid-solution system Bi$_{2-x}$Sb$_x$Te$_{3-y}$Se$_y$ (BSTS) [Fig. 1(b)] provides an excellent platform for studying genuine TIs, because the bulk crystal is highly insulating at certain compositions owing to the reduction of defects and the compensation of defect-induced carriers \cite{RenPRB2010, Cava2012, TaskinPRL, RenPRB2011, TaskinPRB}. The discovery of the BSTS system has stimulated the investigations of electronic structure with various spectroscopies such as angle-resolved photoemission spectroscopy (ARPES) and scanning tunneling microscopy (STM), leading to direct observation of the tunable spin-polarized Dirac carriers \cite{ArakaneNC, HiroshimaBTSSpin, HasanBTS} and the suppressed backscattering channel due to spin-momentum locking \cite{KomoriPRL, STMHiroshima, STMKo}.
  
   After establishing the basic electronic structure of BSTS \cite{ArakaneNC, HiroshimaBTSSpin, HasanBTS, KomoriPRL, STMHiroshima, STMKo}, the next important step is to utilize the novel Dirac-carrier properties such as the spin-momentum locking in actual spintronic devices. An important consequence of the spin-momentum locking is the spin polarization induced by charge current and {\it vice versa}. In principle, while the net spin polarization in the Dirac-cone state is zero at equilibrium, injection of charge current generates a finite spin polarization due to the drift of Fermi surface in the momentum space. Conversely, the injection of spins into the Dirac-cone state induces electromotive force to drive charge currents. Such a spin-electricity conversion has been recently investigated by spin-transfer torque \cite{MellnikNature2014, FanNM2014, WangPRL2015}, spin pumping \cite{ShiomiPRL2014, JamaliNL2015, DoraniPRB2014}, and all-electrical measurements of the charge-current-induced spin polarization \cite{LiNN2014, AndoNL2014, DankertNL2015, TangNL2014, LiuPRB2015, LeePRB2015, TianSR2015, YangArXiv2016}.
   
   While such studies have opened a pathway toward application of TIs in spintronics, the performance of devices so far achieved is not as high as one would expect from theory \cite{ShiomiPRL2014}. This is largely related to the TI-metal contact necessary for operating the electrical circuits and devices; for example, electrodes are always required for electrical measurements, and ferromagnetic metals are used as a spin detector in the measurements of current-induced spin polarizations \cite{LiNN2014, AndoNL2014, DankertNL2015, TangNL2014, LiuPRB2015, LeePRB2015, TianSR2015, YangArXiv2016}. It is thus essential to well characterize the TI-metal interface to integrate bulk insulating TIs into efficient devices. In this regard, it is of great importance to know the work function of bulk insulating TIs, since the band bending caused by a difference in the work functions between a TI and a metal largely influences the interface property and thereby plays a crucial role for device operation. However, the work function of TIs has been reported only for limited prototypical TIs \cite{WorkfuncBi2Te3, WorkfuncBi2Se3, WorkfuncSb2Te3}.
   
    In this Letter, we report our measurements of the work function of BSTS at various bulk-insulating compositions, determined precisely by high-resolution photoemission spectroscopy (PES). We observed that the work function of BSTS systematically increases from 4.95 to 5.20 eV upon increasing $x$ from 0.0 to 1.0, consistent with the band diagram obtained from the ARPES study \cite{ArakaneNC}. We also determined the work function of other TIs such as Pb- and Tl-based solid solutions, and compared them with those of BSTS. 
 
   Single crystals of Bi$_{2-x}$Sb$_x$Te$_{3-y}$Se$_y$ were grown by sealing a stoichiometric amount of high-purity elements with various $x$ and $y$ values in an evacuated quartz tube and melting them at 850 $^{\circ}$C for 48 h with intermittent shaking, followed by cooling slowly to 550 $^{\circ}$C and annealing at that temperature for 4 days. Details of the sample preparation were described elsewhere \cite{RenPRB2011, ArakaneNC}. We measured BSTS samples of ($x$, $y$) = (0.0, 1.0), (0.25, 1.15) and (1.0, 2.0). Hereafter we label these samples by simply referring to the $x$ value only. PES measurements have been performed with a Scienta SES-200 electron analyzer with a high-flux He discharge lamp and a toroidal grating monochromator at Tohoku University. During the measurements, a bias voltage was applied to the sample with respect to the electron analyzer with a battery. The He I$\alpha$ ($h\nu$ = 21.218 eV) line was used to excite photoelectrons. Samples were kept at room temperature during the measurements. The energy resolution was set at 20 meV. A shiny mirror-like surface of samples for PES measurements was obtained {\it in-situ} by cleaving.

\begin{figure}
 \includegraphics[width=3.4in]{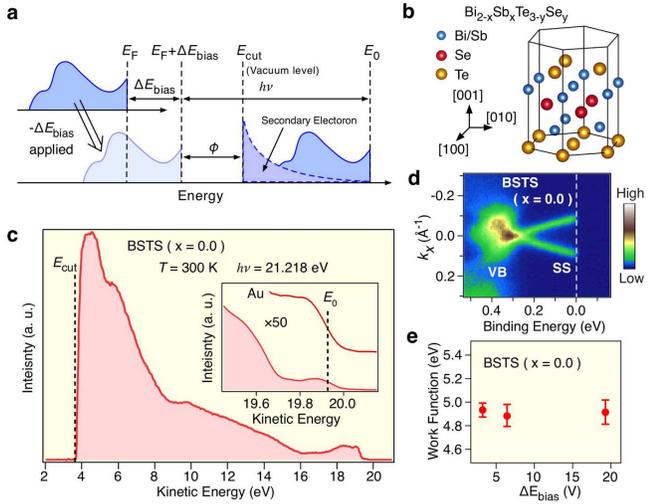}
\caption{(color online). (a) Schematic energy diagram of the photoemission process with applied bias voltage (${\Delta}E_{\rm bias}$). Shaded regions at left- and right-hand sides correspond to the density of states (DOS) and PES spectrum, respectively. $E_0$ and $E_{\rm cut}$ are the kinetic energy of photoelectrons at the Fermi edge and the low-energy cut-off, respectively. $h\nu$ and $\Phi$ are the photon energy and the sample's work function, respectively. (b) Crystal structure of BSTS. (c) PES spectrum of BSTS for $x$ = 0.0 in a wide energy region covering $E_{\rm cut}$ and $E_0$. The intensity is integrated over the momentum window of 0.4 \AA$^{-1}$ centered at the $\bar{\Gamma}$ point. Inset shows an expansion of the PES spectrum around $E_0$, highlighting the existence of a Fermi-edge cut-off. The Fermi edge of gold is also shown as a reference. (d) ARPES intensity plot around the $\bar{\Gamma}$ point as a function of binding energy and wave vector for $x$ = 0.0 (Ref. \onlinecite{ArakaneNC}). (e) Bias-voltage dependence of the work function for $x$ = 0.0.}
\end{figure}

  The work function of materials can be determined by various experimental techniques, such as photon-energy-dependent photocurrent measurement, Kelvin-prove force microscopy (KPM), photoelectron microscopy (PEEM), PES with applying bias voltage, and two-photon PES. Among these, we choose PES with bias voltage, since this technique is capable of determining an intrinsic work-function ($\Phi$) free from the surface contamination, since the measurement is done with a clean surface under ultrahigh vacuum.  As shown in Fig. 1(a), a bias voltage of a few electron volt (${\Delta}E_{\rm bias}$) is applied between the sample and the electron analyzer. When the sum of the bias voltage and the sample's work function exceeds the analyzer's work function $\Phi_{\rm ana}$(i.e. ${\Delta}E_{\rm bias} + \Phi >  \Phi_{\rm ana}$), the low-energy cut-off point in the PES spectrum is directly related to the sample's work function. As shown in Fig. 1(a), the work function is estimated with the energy conservation law, $h\nu = (E_0 - E_{\rm cut}) + \Phi$, where $E_0$ represents the kinetic energy of photoelectrons at the Fermi edge. Since $h\nu$ is fixed at 21.218 eV, one can estimate the work function by simply measuring the energy interval between the Fermi edge and the low-energy cut-off $(E_0 - E_{\rm cut})$ in the PES spectrum.
  
 \begin{figure}
 \includegraphics[width=3.4in]{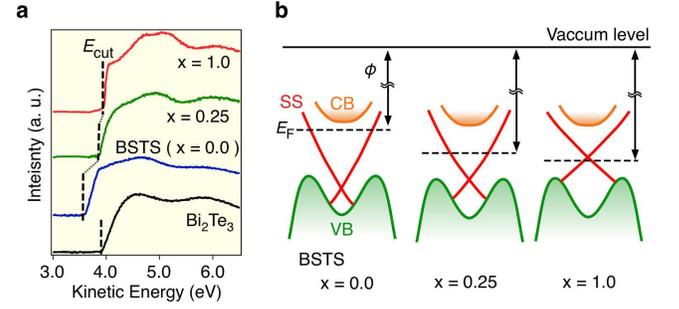}
\caption{(color online). (a) Comparison of PES spectrum around $E_{\rm cut}$ for BSTS with $x$ = 0.0, 0.25, and 1.0. PES spectrum for Bi$_2$Te$_3$ ($n$-type) is also shown for comparison. (b) Band diagram of BSTS for $x$ = 0.0, 0.25, and 1.0. CB, VB, and SS denote the conduction band, the valence band, and the surface state, respectively.}
\end{figure}

  Next, we demonstrate how to determine the work function with an actual sample. Figure 1(c) shows the PES spectrum of BSTS for $x$ = 0.0 measured with ${\Delta}E_{\rm bias}$ = 3.0 eV. The PES spectrum was obtained by integrating the spectral intensity over the momentum window of $\sim$ 0.4 \AA$^{-1}$ centered at the $\bar{\Gamma}$ point, and therefore it represents a partially angle-integrated spectrum. In Fig. 1(c), one can recognize a sharp cut-off in the PES spectrum at low kinetic energy ($E_{\rm cut}$), together with a Fermi-edge cut-off with a much weaker intensity at $E_0$ (see inset). This Fermi edge arises from the Dirac-cone surface state, showing a V-shaped energy dispersion \cite{ArakaneNC} as displayed in Fig. 1(d). We have estimated the work function of this sample to be 4.95 eV, as $\Phi$ (4.95) = $h\nu$ (21.218) - $E_0$ (19.93) + $E_{\rm cut}$ (3.65). As seen in Fig. 1(e), this value is independent of the bias voltage within an experimental accuracy, suggesting that the obtained value of the work function is intrinsic.
  
\begin{table*}
\caption{\label{tab:table1}Work function of various TIs obtained by the present PES experiment. Data for typical metals used for contacting TIs [Cu, Ag, Ti, and Au (Ref. \onlinecite{HerbertJAP1977})] and ferromagnetic metals [Fe (Ref. \onlinecite{HerbertJAP1977}), CoFeB\cite{UhrmannJAP2008}, and permalloy\cite{SaitoVacuum1981}] are also shown for comparison.}
\begin{ruledtabular}
\begin{tabular}{lcccccccc}
 TI Materials & \shortstack{BSTS\\($x$=0.0)} & \shortstack{BSTS\\($x$=0.25)} & \shortstack{BSTS\\($x$=1.0)} & Bi$_2$Te$_3$ & Bi$_2$Se$_3$  & PBST & TBS & \shortstack{TBST\\(Sb 60\%)}\\
\hline
Work function (eV) & 4.95 & 5.15 & 5.20 & 5.25 & 5.60 & 4.80 & 4.70 & 4.85\
\end{tabular}
\vspace{0.3cm}
\begin{tabular}{lccccccc}
Metals & Cu & Ag & Ti & Au & Fe & CoFeB & permalloy\\
\hline
Work function (eV) & 4.65 & 4.26 & 4.33 & 5.1 & 4.5 & 4.8 & 4.83\
\end{tabular}
\end{ruledtabular}
\end{table*}
  
  Next we determined the compositional dependence of the work function. Figure 2(a) shows PES spectra in the energy region around $E_{\rm cut}$ for three different samples with $x$ = 0, 0.25, and 1.0, compared with that for $n$-type Bi$_2$Te$_3$. One can recognize that $E_{\rm cut}$ is located at $\sim$ 3.7 eV for $x$ = 0.0, gradually moves toward higher kinetic energy on increasing $x$, and finally reaches $\sim$ 4.0 eV at $x$ = 1.0. Correspondingly, the work function gradually increases from 4.95 to 5.20 eV with increasing $x$, as shown in Table 1. Such a systematic change in the work function is well correlated with the change in the near-$E_F$ electronic states, as schematically shown in Fig. 2(b). We reported in our previous ARPES study \cite{ArakaneNC} that the $x$ = 0.0 sample shows an $n$-type Dirac cone with the chemical potential lying slightly below the bulk conduction band (CB). On increasing $x$, the chemical potential shifts downward, leading to the $p$-type Dirac carriers with the chemical potential closer to the bulk valence band (VB) at $x$ = 1.0. It is thus expected that the downward chemical-potential shift triggers an increase in the work function. Quantitatively, the change in the work-function value (5.20 - 4.95 = 0.25 eV) is in good agreement with the chemical-potential shift ($\sim$ 0.3 eV) \cite{ArakaneNC}. This suggests that the CB bottom is always pinned at the same position with respect to the vacuum level, indicating the $x$-independent nature of electron affinity.

In Table I, we summarize the work-function value for various TIs obtained in the present study. One can see that the work function of BSTS (4.95-5.20 eV) is smaller than those of prototypical bulk-metallic TIs such as Bi$_2$Te$_3$ (5.25 eV) and Bi$_2$Se$_3$ (5.60 eV). On the other hand, the work function of BSTS is larger than those of other solid-solution TI systems such as Pb(Bi$_{0.6}$Sb$_{0.4}$)$_2$Te$_4$ (PBST) and TlBi$_{0.4}$Sb$_{0.6}$Te$_{2}$ (TBST; this material was recently found to show a bulk-insulating behavior \cite{ChiPRB}). It is also remarked that TlBiSe$_2$ (TBS) has the smallest work function (4.70 eV) among TI samples studied here. In Table I, we also list the work function of nonmagnetic and ferromagnetic metals which are typically used as electrodes or spin injectors/detectors by contacting with TIs. One finds in Table I that copper (Cu, 4.65 eV)\cite{HerbertJAP1977}, silver (Ag, 4.26 eV)\cite{HerbertJAP1977}, and titanium (Ti, 4.33 eV)\cite{HerbertJAP1977} have a smaller work function than those of BSTS, while the work function of gold (Au, 5.1 eV)\cite{HerbertJAP1977} is comparable to BSTS. This result suggests that Cu, Ag and Ti are more useful than Au for obtaining an Ohmic contact with TIs. Ferromagnetic metals, such as iron (Fe, 4.5 eV)\cite{HerbertJAP1977}, CoFeB (4.8 eV)\cite{UhrmannJAP2008} and permalloy (4.83 eV)\cite{SaitoVacuum1981}, commonly have a smaller work function, suggesting that the contact between TIs and these ferromagnets gives rise to the band bending of 0.15-0.7 eV at the interface. Such a band bending due to the work-function difference should be properly taken into account for developing and designing TI-based spintronic devices \cite{YangArXiv2016}.

\begin{acknowledgements}
 We thank K. Fukutani and C. X. Trang for their assistance in the PES experiments. We also thank K. Eto, M. Novak, Z. Ren and Z. Wang for their help in crystal growths. This work was supported by MEXT of Japan (Innovative Area ``Topological Materials Science", 15H05853), and JSPS (KAKENHI 15H02105, 26287071, 25287079, 25220708).
\end{acknowledgements}


\bibliographystyle{prsty}

\end{document}